\documentclass[prd,superscriptaddress, nofootinbib]{revtex4}
\usepackage{graphicx}% Include figure files
\usepackage{dcolumn}% Align table columns on decimal point
\usepackage{slashbox,multirow}
\usepackage{amsmath,amsthm,amssymb}
\usepackage{cases}
\begin{document}
%

%\preprint{APS/123-QED}

\title{Anomalously small wave tails  in higher dimensions}

\author{Piotr Bizo\'n}
\affiliation{M. Smoluchowski Institute of Physics, Jagiellonian
University, Krak\'ow, Poland}
\author{Tadeusz Chmaj}
\affiliation{H. Niewodniczanski Institute of Nuclear
   Physics, Polish Academy of Sciences,  Krak\'ow, Poland}
   \affiliation{Cracow University of Technology, Krak\'ow,
    Poland}
\author{Andrzej Rostworowski}
\affiliation{M. Smoluchowski Institute of Physics, Jagiellonian
University, Krak\'ow, Poland}
\date{\today}
\begin{abstract}
We consider the late-time tails of spherical waves propagating on
even-dimensional Minkowski spacetime  under the influence of a long
range radial potential. We show that in six and higher even
dimensions there exist exceptional potentials for which  the tail
has an anomalously small amplitude and fast decay. Along the way we
 clarify and amend some confounding arguments and statements in the literature of
 the subject.

\end{abstract}

\maketitle

\section{Introduction}
It is well known that sharp propagation of free waves along light
cones in even-dimensional flat spacetimes, known as Huygens'
property, is blurred by the presence of a potential. Physically, the
spreading of waves inside the light cone is caused by the
backscattering off the potential. If the potential falls off
exponentially or faster at spatial infinity, then the backscattered
waves decay exponentially in time, while the long range potentials
with an algebraic fall-off give rise to tails which decay
polynomially in $1/t$. The precise description of these tails is an
important issue in scattering theory. There are two main approaches
to this problem in the literature. On the one hand, there are
mathematical results in the form of various decay estimates. These
results are rigorous, however they rarely give optimal decay rates
inside the light cone and provide very poor information about the
amplitudes of tails. The notable exception is the work of Strauss
and Tsutaya \cite{st} (recently strengthened  by Szpak \cite{sz})
where the optimal pointwise decay estimate for the tail was proved
in four dimensions. Unfortunately, to the best of our knowledge,
there is no analogous result in higher dimensions.

On the other hand, there are non-rigorous results in the physics
literature based on perturbation theory. The most complete work in
this category was done by  Ching \emph{et al.} \cite{ching} who
derived first-order approximations of the tails  for radial
potentials. Although these results were originally formulated for
partial waves in four dimensions, they can be easily translated to
spherical waves in higher dimensions. Ching \emph{et al.} noticed
that there are exceptional potentials for which the first-order tail
vanishes, however they did not pursue their analysis to the second
order, apart from giving some dimensional arguments. The main
purpose of this paper is to analyze the tails for such exceptional
potentials in more detail.

One of the physical motivations behind our work stems from the fact
that this kind of potentials arise in the study of linearized
perturbations of higher even-dimensional Schwarzschild black holes.
The behavior of tails on the Schwarzschild background  in well known
in four dimensions (see \cite{p}, \cite{l}, \cite{ching},
\cite{gpp}, \cite{b}, \cite{dr}), but not in higher even dimensions
(despite statements to the contrary in the literature \cite{car}).
Although our analysis is restricted to the flat background, it sheds
some light on the problem of tails on the black hole background
because the properties of tails are to some extent independent of
what happens in the central region.

 The rest of the paper is organized as follows. In section~2 we
 construct the iterative scheme for the perturbation expansion
of a spherically symmetric solution of the linear wave equation with
a potential.
  This scheme is applied in
 section~3 to derive the first and second-order approximations of the tails for radial potentials
which fall off as pure inverse-power at infinity.
 In section~4 we discuss
 the modifications caused by  subleading terms in the potential. Section~5 contains numerical
 evidence
  confirming the analytic formulae from sections~3 and 4. Finally, in section~6
  we give a heuristic argument  to predict the behavior of tails outside
  Schwarzschild
 black holes in higher even dimensions. Technical details of most calculations are given in
 the appendix.

Throughout the paper we use the succinct notation and summation
technics from the excellent book by Graham \emph{et al.} \cite{gkp}.
In particular, we shall frequently use the following abbreviations
\begin{eqnarray}
x^{\underline{0}} := 1, &\qquad& x^{\underline{k}} := x \cdot (x-1)
\cdot \dots \cdot (x-(k-1)), \quad k>0\,,
\\
x^{\overline{0}} := 1, &\qquad& x^{\overline{k}} := x \cdot (x+1)
\cdot \dots \cdot (x+(k-1)), \quad k>0\,.
\end{eqnarray}

\section{Iterative scheme}
We consider the wave equation with a potential in even-dimensional
Minkowski spacetime $R^{d+1}$
\begin{equation}\label{eqm}
  \partial_t^2 \phi -\Delta \phi + \lambda V \phi =0\,.
\end{equation}
 The prefactor
$\lambda$ is introduced for convenience - throughout the paper we
assume that $\lambda$ is small which allows us to use it as the
perturbation parameter. The precise assumptions about the fall-off
of the potential will be formulated below. We restrict attention to
spherical symmetry, i.e., we assume that $\phi=\phi(t,r)$ and
$V=V(r)$. Then, equation (\ref{eqm})  becomes
\begin{equation}\label{eqs}
  \mathcal{L} \phi + \lambda V(r) \phi =0\,,\qquad \mathcal{L}
   :=\partial_t^2  -\partial_r^2  -\frac{d-1}{r} \partial_r\,.
\end{equation}
 We are interested in the late-time behavior
of $\phi(t,r)$ for smooth compactly supported (or exponentially
localized) initial data.
\begin{equation}\label{id}
    \phi(0,r)=f(r),\qquad \partial_t \phi(0,r)=g(r)\,.
\end{equation}
 To determine the asymptotic behavior of solutions
we define the perturbative expansion (Born series)
\begin{equation}\label{pert}
    \phi=\sum_{n=0} \lambda^n \phi_n\,,
\end{equation}
where $\phi_0$ satisfies initial data (\ref{id}) and all $\phi_n$
with $n>0$ have zero initial data. Substituting this expansion into
equation (\ref{eqs}) we get the iterative scheme
\begin{equation}\label{scheme}
  \mathcal{L} \phi_n = -V \phi_{n-1}\,, \qquad \phi_{-1}=0\,,
  \end{equation}
  which can be solved recursively.  The zeroth-order solution is given by the general
regular solution of the free radial wave equation  which is a
superposition of outgoing and ingoing waves \cite{k}
\begin{equation}\label{f0}
    \phi_0(t,r)=\phi_0^{ret}(t,r)+\phi_0^{adv}(t,r)\,,
\end{equation}
where
\begin{equation}\label{f1}
    \phi_0^{ret}(t,r)= \frac{1}{r^{l+1}}\,\sum_{k=0}^{l}  \frac {(2l-k)!} {k!(l-k)!} \frac
{a^{(k)}(u)}{(v-u)^{l-k}}\,, \qquad   \phi_0^{adv}(t,r) =
\frac{1}{r^{l+1}}\,\sum_{k=0} ^{l} (-1)^{k+1} \frac {(2l-k)!}
{k!(l-k)!} \frac {a^{(k)}(v)}{(v-u)^{l-k}}\,,
\end{equation}
and $u=t-r$, $v=t+r$ are the retarded and advanced times,
respectively. Here and in the following, instead of $d$, we use the
index $l$ defined by $d=2l+3$ (remember that we consider only
\textit{odd} space dimensions $d$).
 Note that for
compactly supported initial data the generating function $a(x)$ can
be chosen to have compact support as well (this condition determines
$a(x)$ uniquely).

 To solve equation (\ref{scheme}) for the higher-order perturbations we use
   the Duhamel representation for the solution of the inhomogeneous
equation $\mathcal{L} \phi = N(t,r)$ with zero initial data
\begin{equation}\label{a2}
    \phi(t,r)= \frac{1}{2 r^{l+1}}
    \int\limits_{0}^{t} d\tau \int\limits_{|t-r-\tau|}^{t+r-\tau} \rho^{l+1} P_l(\mu)
 N(\tau,\rho)  d\rho\,,
\end{equation}
where $P_l(\mu)$ are Legendre polynomials of degree $l$ and
$\mu=(r^2+\rho^2-(t-\tau)^2)/2r\rho$ (note that $-1\leq \mu \leq 1$
within the integration range). This formula can be readily obtained
by integrating out the angular variables in the standard formula
$\phi=G^{ret} * N$ where $G^{ret}(t,x)=(2\pi^{l+1})^{-1}\Theta(t)
\delta^{(l)}(t^2-|x|^2)$ is the retarded Green's function of the
wave operator  in $d+1$ dimensions (see, for example, \cite{ls}).

It is convenient to express (\ref{a2}) in terms of null coordinates
$\eta=\tau-\rho$ and $\xi=\tau+\rho$
\begin{equation}\label{duh1}
    \phi(t,r)= \frac{1}{2^{l+3} r^{l+1}}
    \int\limits_{|t-r|}^{t+r} d\xi \int\limits_{-\xi}^{t-r} (\xi-\eta)^{l+1} P_l(\mu)
    N(\eta,\xi)  d\eta\,,
\end{equation}
where now $ \mu=(r^2+(\xi-t)(t-\eta))/r(\xi-\eta)$.
Using this representation we can rewrite the iterative scheme
(\ref{scheme}) in the integral form
\begin{equation}\label{iter}
    \phi_{n}(t,r)= -\frac{1}{2^{l+3} r^{l+1}}
    \int\limits_{|t-r|}^{t+r} d\xi  \int\limits_{-\xi}^{t-r}
   (\xi-\eta)^{l+1} P_l(\mu) V(\rho(\eta,\xi))\phi_{n-1}(\eta,\xi)
    d\eta\,.
\end{equation}
This "master" equation will be applied below to evaluate the first
two iterates for a special class of potentials. It
  is natural to expect that for sufficiently small
$\lambda$ these iterates provide good approximations of the true
solution.

\section{Pure inverse-power potentials at infinity}
In this section we consider the simple case (below referred to as
type~I) when the potential is \emph{exactly} $V(r)= r^{-\alpha}$ for
$r$ greater than some $r_0>0$. We assume that $\alpha>2$. The
modifications caused by subleading corrections to the pure
inverse-power decay of the potential will be discussed in section~4.
\subsection{Generic case}
 We wish to evaluate the  first iterate $\phi_1(t,r)$  near
timelike infinity, i.e, for $r=const$ and $t\rightarrow \infty$.
Thanks to the fact that $\phi_0(\eta,\xi)$ has compact support we
may interchange the order of integration  in (\ref{iter}) and drop
the advanced part of $\phi_0(\eta,\xi)$ to obtain
\begin{equation}\label{iter1}
    \phi_{1}(t,r) = -\frac{2^{\alpha}}{2^{l+3} r^{l+1}}
    \int\limits_{-\infty}^{\infty} d\eta  \int\limits_{t-r}^{t+r} (\xi-\eta)^{l+1-\alpha}
    P_l(\mu)  \phi_0^{ret}(\eta,\xi)
    d\xi\,,
\end{equation}
where we have substituted $V= 2^{\alpha} (\xi-\eta)^{-\alpha}$.
Plugging (\ref{f1}) into (\ref{iter1}), after a long calculation
(see appendix A for the technical details), we get
\begin{eqnarray}
 \phi_1(t,r) &=& - 2^{\alpha+3l-1} \left( \frac {\alpha-3} {2}
\right)^{\underline{l}} \left( \frac {\alpha} {2}
\right)^{\overline{l}} \int \limits_{-\infty}^{+\infty} d\eta \,
 a(\eta) \, \frac {(t-\eta)^{\alpha-2}} {\left[ (t-\eta)^2 - r^2 \right]^{\alpha-1+l}}
\nonumber\\
&\times& \sum_{0 \leq n \leq \lfloor (\alpha-2) / 2 \rfloor} (-1)^{n} \frac {2^{2n}(l+n)!} {n! (2l+2n+1)!}
\left( - \frac {\alpha-2} {2} - l - 1 \right)^{\underline{n}} \left( \frac {\alpha-1} {2} - l - 1 \right)^{\underline{n}}
\nonumber\\
&\times& \sum_{n \leq m \leq n+l} (-1)^{m} \left( \begin{array}{c} l \\ m-n \end{array} \right)
 \frac {\left( - \frac {\alpha} {2} + 1 \right)^{\overline{m}}} {\left( \frac {\alpha} {2} \right)^{\overline{m}}} \left( \frac {r} {t-\eta} \right)^{2m}.
\label{tailf1}
\end{eqnarray}
Asymptotic expansion of (\ref{tailf1}) near timelike infinity yields
the following first-order approximation of the tail
\begin{equation}\label{tail1}
   \phi(t,r) \approx \lambda \phi_1(t,r) = \lambda \, \frac {C(l,\alpha)} {t^{\alpha+2l}} \left[ A + (\alpha+2l)\frac{B}{t}  \, + \mathcal{O} \left( \frac{1}{t^2}
\right) \right] \, ,
    \end{equation}
    where
    \begin{equation}\label{B}
        C(l,\alpha) = - \frac {2^{\alpha+2l-1}}{(2l+1)!!} \left( \frac {\alpha-3} {2} \right)^{\underline{l}}
        \left( \frac {\alpha} {2} \right)^{\overline{l}}\,,
    \end{equation}
    and
    \begin{equation}\label{ab}
        A=\int \limits_{-\infty}^{+\infty} a(\eta)\, d\eta\,,\qquad B=\int \limits_{-\infty}^{+\infty}
        a(\eta)\,\eta\,d\eta\,.
    \end{equation}
In general $A\neq 0$ and the tail decays as $t^{-\alpha-2l}$,
however there are nongeneric initial data for which $A=0$ and
    then the
    tail  decays as $t^{-\alpha-2l-1}$; in particular this happens for  time
    symmetric initial data for which $a(x)$ is an odd function.
\vskip 0.2cm \noindent \emph{Remark~1.} It is easy to check that if
the function $\phi(t,r)$ satisfies equation (\ref{eqs}),  then the
function $\psi=r^{l+1}\phi$ satisfies the radial wave equation for
the $l$th multipole
\begin{equation}\label{a3}
(\partial_t^2-\partial_r^2+l(l+1)/r^2)\psi + \lambda V(r)\psi =0\,.
\end{equation}
The late-time tails for this equation were studied
 by Ching \emph{et al.} \cite{ching} who derived the formula equivalent to (\ref{tail1})
  via the Fourier transform methods.
\subsection{Exceptional case}
It follows from (\ref{tail1}) that if $\alpha$ is an odd integer
satisfying $3\leq \alpha \leq 2l+1$, then $\phi_1(t,r)$ vanishes
identically
  due to factor $\left( \frac {\alpha-3} {2} \right)^{\underline{l}}$ in (\ref{B}) and there is no (polynomial) tail whatsoever
in the first order. Thus, in order to compute the tail in this
exceptional case we need to go the second order of the perturbation
expansion.

Using (\ref{iter}) and proceeding as above we get the
 second iterate
\begin{equation}\label{iter2}
    \phi_{2}(t,r) = -\frac{2^{\alpha}}{2^{l+3} r^{l+1}}
    \int\limits_{-\infty}^{\infty} d\eta  \int\limits_{t-r}^{t+r}
    (\xi-\eta)^{l+1-\alpha} P_l(\mu)  \phi_1^{ret}(\eta,\xi)
    d\xi\,,
\end{equation}
where $\phi_1^{ret}$ is the outgoing solution of the inhomogeneous
equation
\begin{equation}\label{NH}
\mathcal{L} \phi_1=-V \phi_0\,.
\end{equation}
In general $\phi_1$ is a sum of the solution of the homogeneous
equation and the particular solution of the inhomogeneous equation.
The homogeneous part has the form (\ref{f1}) (with a different
generating function than $a$, but still compactly supported), thus
for the same reason as above it gives no contribution to the tail.
The particular solution of the inhomogeneous equation (\ref{NH})
reads
\begin{equation}
\label{f1nul} \phi_l^{NH} = \frac {1}{2 (\alpha-1) r^{\alpha +l}}
\sum_{q=0}^{l-\alpha/2+1/2} (l-\alpha/2+1/2)^{\underline{q}} \;
\frac {2^q \left(\alpha/2 \right)^{\overline{q}}}
{\alpha^{\overline{q}}} \frac {\phi_{l-1-q}^{H}} {r^q} \, ,
\end{equation}
where $\phi_{l-1-q}^{H}$ denotes the solution of the homogeneous
equation with $d=2(l-1-q)+3$ and the same generating function $a$ as
in $\phi_0$ (see (\ref{f1})). The  formula (\ref{f1nul}) can be
easily derived by the method of undetermined coefficients (we
emphasize that this formula is valid \emph{only} for odd $\alpha$
satisfying $3\leq \alpha \leq 2l+1$).
Substituting (\ref{f1nul}) into (\ref{iter2}), after a long
calculation (see appendix A for the technical details), we obtain
the following asymptotic behavior near timelike infinity
\begin{equation}
\label{tail2}
  \phi(t,r) \approx \lambda^2 \, \phi_2(t,r) = \lambda^2 \frac {D(l,\alpha)} {t^{2(\alpha+l-1)}}
  \left[ A + 2(\alpha+l-1)\frac{B}{t}  + \mathcal{O} \left( \frac{1}{t^2} \right) \right]\,,
    \end{equation}
    where the coefficients $A$ and $B$ are defined in (\ref{ab}) and
    \begin{equation}\label{c2}
        D(l,\alpha)= \frac {2^{2(\alpha+l-2)}}{(2l+1)!!} \cdot \frac{(2\alpha-3)}{2(\alpha-1)}
        \left( \alpha - \frac {5} {2} \right)^{\underline{l-1}} \left( \alpha - 2 + l
        \right)^{\underline{l-1}}\,
         F\left( \left.
          \begin{array}{c} -l+\alpha/2-1/2,\, \alpha/2,\, 2\alpha-2,\, 1 \\ \alpha,\, \alpha,\,
          \alpha - l-1/2  \end{array} \right| 1
          \right)\,.
    \end{equation}
     Here $F$ stands for the generalized hypergeometric function
     \begin{equation}
     \label{F}
F\left( \left.
\begin{array}{c}
a_1,\, \dots,\, a_m
\\
b_1,\, \dots,\, b_n
\end{array}
\right| z \right) = \sum _{k \geq 0} \frac {a_1^{\overline{k}},\,
\dots,\, a_m^{\overline{k}}}
 {b_1^{\overline{k}},\, \dots,\, b_n^{\overline{k}}}
 \frac{z^k}{k!}\,.
\end{equation}
\renewcommand{\arraystretch}{1.5}
\setlength{\abovecaptionskip}{6pt}   % 0.5cm as an example
\setlength{\belowcaptionskip}{6pt}   % 0.5cm as an example
\setlength{\tabcolsep}{1.2em}
\setlength{\baselineskip}{16.0pt}    % 16 pt usual spacing between lines
\begin{table}[h]
  \centering
  \caption{The first few coefficients $D(l,\alpha)$}\label{table 1}
\begin{tabular}{|c||*{4}{c|}}\hline
\backslashbox{$l$}{$\alpha$} & \makebox[3em]{$3$} &
\makebox[3em]{$5$} & \makebox[3em]{$7$} & \makebox[3em]{$9$}
\\
\hline \hline $1$ & $4$ & & &
\\
\hline $2$ & $-8/5$ & $2240/3$ & &
\\
\hline $3$ & $96/35$ & $1792$ & $2523136/5$ &
\\
\hline $4$ & $-64/7$ & $-17920/9$ & $16580608/5$ & $4638965760/7$
\\
\hline
\end{tabular}
\end{table}

\noindent We remark that the behavior $\mathcal{O}(\lambda^2)\,
t^{-2(l+\alpha-1)}$ of the tail (\ref{tail2}) was conjectured before
by Ching \emph{et al.} \cite{ching} on the basis of dimensional
analysis.
\section{General polynomially decaying potentials}
In this section we analyze  how the presence of subleading
corrections to the pure inverse-power asymptotic  behavior of the
potential affects the results obtained in section~3. We restrict
ourselves to the most interesting and common case (below referred to
as type~II) when near infinity
 \begin{equation}\label{case2}
        V(r) = \frac{1}{r^{\alpha}} \left(1+\frac{\beta}{r^{\gamma}}\right)
        +o\left(\frac{1}{r^{\alpha+\gamma}}\right),\,\qquad
        \gamma>0\,.
    \end{equation}
    If $C(\alpha,l)\neq 0$, then the dominant behavior of the tail
   is of course the same as in (\ref{tail1}):
\begin{equation}\label{gtail1}
\phi(t,r) \sim \lambda \, A  \, C(l,\alpha)\,
   t^{-(\alpha+2l)}\,.
\end{equation}
However, in the exceptional case, when $C(\alpha,l)=0$, the
situation is more delicate.
 As we showed above, in
this case there is the second-order contribution to the tail given
by  (\ref{tail2})
\begin{equation}\label{gtail2}
\phi_2(t,r) \sim  A  \, D(l,\alpha)\,
   t^{-2(\alpha+l-1)}\,.
\end{equation}
In contrast to  the type~I case where the first-order tail vanishes
identically, in the type~II case  the subleading term in the
potential produces the first-order contribution which is given by
(\ref{tail1}) with $\alpha$ replaced by $\alpha+\gamma$:
\begin{equation}\label{tailf1p}
   \phi_1(t,r) \sim \beta \, A\, C(l,\alpha+\gamma)\,
   t^{-(\alpha+\gamma+2l)}\,,
    \end{equation}
   assuming that $\alpha+\gamma$ is not an odd integer $\leq
    d-2$ (otherwise one has to repeat the analysis for the next subleading term in the potential).

 Now, comparing the decay rates in (\ref{gtail2}) and
(\ref{tailf1p}) we conclude that the leading asymptotics of the tail
is given by the first-order term $\lambda \phi_1(t,r)$  if
$\gamma\leq \alpha-2$ (we call it subtype~IIa), but otherwise,
\emph{i.e.} for $\gamma
> \alpha-2$ (subtype~IIb), the second-order term $\lambda^2 \phi_2(t,r)$ is
dominant for $t\rightarrow \infty$. \vskip 0.2cm \noindent
\emph{Remark~2.} In the context of equation (\ref{a3}) a formula
analogous to (\ref{tailf1p}) was obtained by Hod who studied tails
in the presence of subleading terms in the potential (see subgroup
IIIb in \cite{hod}). However, Hod's analysis, restricted to the
first-order approximation, was inconclusive because, as we just have
shown, without the second-order formula (\ref{gtail2}) one is  not
in position to make assertions about the dominant behavior of the
tail. \vskip 0.2cm
\section{Numerics}
In order to verify the above analytic predictions we solved
numerically the initial value problem (4-5) for various potentials
and initial data.  Our numerical algorithm is based on the method of
lines with finite differencing in space and explicit fourth-order
accurate Runge-Kutta time stepping.  As was pointed out in
\cite{ching}, a reliable numerical computation of tails requires
high-order finite-difference schemes, since otherwise the ghost
potentials generated by discretization errors produce artificial
tails which might mask the genuine behavior. The minimal order of
spatial finite-difference operators depends on the fall-off of the
potential -- for the cases presented below the fourth-order accuracy
was sufficient, but for the faster decaying potentials a
higher-order accuracy is needed. To eliminate high-frequency
numerical instabilities we added a small amount of Kreiss-Oliger
artificial dissipation  All computations were performed using
quadruple precision which was essential in suppressing round-off
errors at late times.

 The numerical results
presented here were produced for initial data of the form
\begin{equation}\label{idn}
    \phi(0,r) = \exp(-r^2),\qquad
   \partial_t \phi(0,r)= \exp(-r^2)\,.
\end{equation}
As follows from (\ref{f1})  the generating function for these data
is
\begin{equation}\label{a}
a(x)=2^{-(l+2)} (1-2x) \exp(-x^2)\,,\quad \mbox{hence}\quad
A=\int_{-\infty}^{+\infty} a(x) dx=\sqrt{\pi}/2^{l+2}\,.
\end{equation}
We considered  the following potentials
\begin{subequations}
 \begin{numcases}{V(r)=}\dfrac{\tanh^{\alpha+2}{r}}{r^{\alpha}} &
\quad \mbox{(type
    I})\\
    \dfrac{\tanh^{\alpha+2}{r}}{r^{\alpha}}\left(1+\dfrac{\tanh^{\gamma}{r}}{r^{\gamma}}\right) & \quad \mbox{(type
    II)}\,,
    \end{numcases}
\end{subequations}
for various values of $\alpha$ and $\gamma$. The regularizing factor
$\tanh(r)$ introduces exponentially decaying corrections to the pure
inverse-power behavior at infinity but such corrections do not
affect the polynomial tails.
The numerical verification of the formulae (\ref{tail1}),
(\ref{tailf1p}), and (\ref{tail2}) is shown in tables II and III.
The observed decay rates agree perfectly with analytic predictions,
while small errors in the amplitudes
%(of the order of a few percent)
are due to (neglected) higher-order terms in the perturbation
expansion.

\begin{table}[h]
  \centering
   \caption{The generic case: numerical verification of the analytic formula (\ref{tail1}) for
   the potential (31a)
    ($\lambda=0.1$)  and initial data
(\ref{idn}). Comparing the second column of this table
(corresponding to $\alpha=3.01$) with the last column of table~III
     one can see the discontinuity of the decay rate at $\alpha=3$ (for $d=5$ and $7$).\vspace{0.3cm}} \label{table 3}
\begin{tabular}{|c c||c|c|c|c|c|c|}
  \hline
& &\multicolumn{2}{c|}{$\alpha=2.5$ }&\multicolumn{2}{c|}{$\alpha=3.01$ }&\multicolumn{2}{c|}{$\alpha=4$}\\
\cline{3-8}
  & & Theory & Numerics & Theory & Numerics & Theory & Numerics\\ \hline \hline
  \multirow{2}{*}
  {$d=3$} & Exponent &  2.5 & 2.499 & 3.01 & 3.009 & 4 & 4.00002 \\
  & Amplitude & -0.1253 & -0.0881 &  -0.1785 & -0.1518 & -0.3545 &  -0.3320 \\ \hline
  \multirow{2}{*}
  {$d=5$} & Exponent  & 4.5 & 4.501 & 5.01 & 5.0101 & 6 & 5.9999
\\
  & Amplitude & 0.0261 &0.0235 & -0.00089  & -0.00085 & -0.2363  &  -0.2318\\ \hline
  \multirow{2}{*}
{$d=7$} & Exponent & 6.5 & 6.501 & 7.01 & 7.01 & 8 & 7.9999   \\
 & Amplitude  & -0.0294& -0.0276 & 0.00089 & 0.00087& 0.1418 & 0.1404 \\
  \hline
\end{tabular}
\end{table}
\begin{table}[h]
  \centering
   \caption{The exceptional case: comparison of analytic and numerical parameters of the tails for  the
   potential (31b) (the first two columns) and (31a) (the third column)
    with $\alpha=3$, $\lambda=0.1$, and initial data
(\ref{idn}).  The analytic results are given by the formula
(\ref{tailf1p}) for the subtype IIa potential, and by the formula
(\ref{gtail2}) for the type I and IIb potentials. Note that although
the dominant tails for the type I and the subtype IIb potentials are
theoretically the same, in the case IIb there is an additional first
order error due to the subdominant term $\mathcal{O}(\lambda)
t^{-(2l+\alpha+\gamma)}$ which accounts for a slight difference in
numerical accuracy between these two cases.\vspace{0.3cm}}
\label{table 3}
\begin{tabular}{|c c||c|c|c|c|c|c|}
  \hline
& &\multicolumn{2}{c|}{$\gamma=1/2$ (subtype IIa)}&\multicolumn{2}{c|}{$\gamma=1.75$ (subtype IIb)}&\multicolumn{2}{c|}{(type I)}\\
\cline{3-8}
  & & Theory & Numerics & Theory & Numerics & Theory & Numerics\\ \hline \hline
  \multirow{2}{*}
  {$d=5$} & Exponent &  5.5 & 5.4993 & 6 & 6.002 & 6 & 6.0000 \\
  & Amplitude &  -0.0731 & -0.0696  &  0.00886 & 0.00862 & 0.00886 &  0.00843 \\ \hline
  \multirow{2}{*}
  {$d=7$} & Exponent  & 7.5 & 7.4998& 8 & 8.0003 & 8& 7.9999\\
  & Amplitude & 0.0603 & 0.0579 &  -0.00177 & -0.00175 & -0.00177  & -0.00172\\ \hline
  \multirow{2}{*}
{$d=9$} & Exponent & 9.5 & 9.4999 & 10 & 9.9957 & 10 & 9.9997   \\
 & Amplitude  & -0.1131 & -0.1115 & 0.00152 & 0.00145 & 0.00152 & 0.00149\\
  \hline
\end{tabular}
\end{table}

\section{Schwarzschild background}
Consider the evolution of the massless scalar field outside the
$d+1$ dimensional Schwarzschild black hole
\begin{equation}\label{sch}
    ds^2=-\left(1-\frac{1}{r^{d-2}}\right) dt^2 +
    \left(1-\frac{1}{r^{d-2}}\right)^{-1}
     dr^2 + r^2
    d\Omega_{d-1}^2\,,
\end{equation}
where $d\Omega_{d-1}^2$ is the round metric on the unit sphere
$S^{d-1}$ and $d\geq 5$ is odd. Here we use units in which  the
horizon radius is at $r=1$. Introducing the tortoise coordinate $x$,
defined by $dr/dx=1-1/r^{d-2}$, and decomposing the scalar field
into multipoles,  one obtains the following reduced wave equation
for the $j$th multipole \cite{ik}
\begin{equation}\label{eqsch}
    \partial_t^2\psi -\partial_x^2 \psi +U(x) \psi=0, \qquad
    U=\left(1-\frac{1}{r^{d-2}}\right) \left(\frac{(2j+d-3)(2j+d-1)}{4r^2}+\frac{(d-1)^2}{4
    r^d}\right)\,.
\end{equation}
Note that (\ref{eqsch}) is the $1+1$ dimensional wave equation  on
the whole axis $-\infty<x<\infty$.
 For large positive $x$ we have
\begin{equation}\label{exp2}
    r =
    x+\frac{1}{d-3}\frac{1}{x^{d-3}}-\frac{d-2}{(2d-5)(d-3)}\frac{1}{x^{2d-5}} +\mathcal{O} \left( \frac{1}{x^{3d-7}}\right)\,,
\end{equation}
which implies that
\begin{equation}\label{expv}
    U(x) =
    \frac{(2j+d-3)(2j+d-1)}{4x^2}+V(x)\,,\qquad V(x)=
    \frac{a}{x^d}+\frac{b}{x^{2d-2}} +\mathcal{O} \left( \frac{1}{x^{3d-4}}\right)   \quad \mbox{as}\quad x\rightarrow
    \infty\,,
\end{equation}
with
\begin{equation}
\label{wsp} a = - \frac {(d-1) j (j+d-2)} {d-3} \qquad \mbox{and}
\qquad b = - \frac {(2d - 3) ((d-3)(d-2)^2(d-1) - 4 j
(j+d-2)(1+d(d-3)))} {4(2d-5)(d-3)^2}\,.
\end{equation}
For large negative $x$ (near the horizon) the potential is
exponentially small, so one expects that the backscattering off the
left edge of the potential can be neglected. If so, the decay rate
(but not the amplitude!) should follow from the analysis of
section~4. Comparing  equation (\ref{eqsch}) for large positive $x$
to equation (\ref{a3}) with the potential (\ref{case2}) and using
(\ref{expv}) we find that $l=j+(d-3)/2$ and the potential $V$ is of
the subtype~IIa with $\alpha=d$ and $\gamma=d-2$. Thus, applying
(\ref{tailf1p}) we get the first-order tail
\begin{equation}\label{tails}
    \psi(t,x) \sim t^{-(2j+3d-5)}\,.
\end{equation}
\emph{Remark~3.} Late-time tails outside higher dimensional
Schwarzschild black holes were studied in \cite{car}, however in the
even-dimensional case the reasoning presented there is not correct,
even though the result agrees with (\ref{tails}). The reason is that
the analysis of \cite{car} is based on the application of Ching
\emph{et al.} conjecture about the decay of the second-order tail
$t^{-(2l+2\alpha-2)}$ which for $l=j+(d-3)/2$ and $\alpha=d$ gives
$t^{-(2j+3d-5)}$. Unfortunately, this conjecture does not apply to
the problem at hand. For $j=0$ this is evident because the leading
term in $V$ (proportional to $x^{-d}$) vanishes (since by
(\ref{wsp}) $a=0$), while the subleading term (proportional to
$x^{-(2d-2)}$) is of generic type. For $j>0$ this follows from the
fact that the potential is of the subtype~IIa. Thus, for all $j\geq
0$ the dominant (first-order) contribution to the tail comes from
the subleading term in the potential. The agreement of the decay
rate obtained in \cite{car} with (\ref{tails}) is accidental and due
to the fact that the subdominant term in (\ref{expv}) (not
considered in \cite{car}) is on a borderline between subtypes IIa
and IIb.

 Admittedly, the  handwaving argument leading to (\ref{tails}) is
far from satisfactory. Unfortunately, we have not been able to carry
over the analysis from sections~2-4 in the case of equation
(\ref{eqsch}). There are two  difficulties in this respect. First,
in contrast to the spherical case, Huygens' principle is not valid
for the free wave equation in $1+1$ dimensions. Second, there is no
natural small parameter in the problem. In the impressive tour de
force work \cite{b} Barack showed how to overcome these difficulties
for a restricted class of initial data in four dimensions. It would
be interesting to generalize Barack's approach to higher
even-dimensional Schwarzschild spacetimes.

\newpage
\vskip 0.3cm \noindent \textbf{Acknowledgments:} PB thanks Nikodem
Szpak for helpful discussions and Leor Barack for clarifying some
details of the paper \cite{b}. AR thanks Prof. Bernd Br\"ugmann for
hospitality in his group at FSU Jena, where a part of this work was
done. This research was supported in part by the MNII grant
1PO3B01229 and grant 189/6. PR UE/2007/7.

\appendix

%%%%%%%%%%%%%%%%%%%%%%%%%%%%%%%%%%%%%%%%%%%%%%%%%%%%%%%%%%%%%%%%%%%%%%%%%%%%%%%%%%%%%%%%%%%%
%%%%%%%%%%%%%%%%%%%%%%%%%%%%%%%%%%%%%%%%%%%%%%%%%%%%%%%%%%%%%%%%%%%%%%%%%%%%%%%%%%%%%%%%%%%%
%%%%%%%%%%%%%%%%%%%%%%%%%%%%%%%%%%%%%%%%%%%%%%%%%%%%%%%%%%%%%%%%%%%%%%%%%%%%%%%%%%%%%%%%%%%%
\section{}
%%%%%%%%%%%%%%%%%%%%%%%%%%%%%%%%%%%%%%%%%%%%%%%%%%%%%%%%%%%%%%%%%%%%%%%%%%%%%%%%%%%%%%%%%%%%
%%%%%%%%%%%%%%%%%%%%%%%%%%%%%%%%%%%%%%%%%%%%%%%%%%%%%%%%%%%%%%%%%%%%%%%%%%%%%%%%%%%%%%%%%%%%
%%%%%%%%%%%%%%%%%%%%%%%%%%%%%%%%%%%%%%%%%%%%%%%%%%%%%%%%%%%%%%%%%%%%%%%%%%%%%%%%%%%%%%%%%%%%
\noindent Throughout the appendix we use the notation of \cite{gkp}
in which the square bracket around a logical expression returns a
value $1$ if the expression is true and a value $0$ if the
expression is false:
\begin{equation}
[condition] = \left\{ \begin{array}{ccl} 1 & \mbox{if} & condition =
\mbox{true} \\ 0 & \mbox{if} & condition = \mbox{false} \end{array}
\right.\nonumber
\end{equation}
In order to derive the asymptotic behavior of the iterates
(\ref{iter1}) and (\ref{iter2}) near timelike infinity (fixed $r$
and $t\rightarrow\infty$) we need to evaluate the following
expression
\begin{equation}
\mathcal{F} (t,r;\,\beta,\,L) = - \frac {2^{\beta}}{4 r^{l+1}}
\sum_{k=0}^L c_{L,k} \, \int \limits_{-\infty}^{+\infty} d\eta \,
\int \limits_{t-r}^{t+r} d\xi \, \frac {P_l (\mu)}
{(\xi-\eta)^{\beta+L-k}} a^{(k)}(\eta), \label{master}
\end{equation}
where
\begin{equation}
c_{L,k} = \frac {(2 L - k)!} {k! (L-k)!}
\end{equation}
and
\begin{equation}
\label{mu} \mu = \frac {(\xi-t)(t-\eta)+r^2}{r(\xi-\eta)}\, .
\end{equation}
From (\ref{f1}) and (\ref{iter1}) we have
\begin{equation}
\phi_{1}(t,r) = \mathcal{F} (t,r;\,\alpha,\,l),
\end{equation}
and from (\ref{iter2}) and  (\ref{f1nul}) we have
\begin{equation}\label{a5}
\phi_{2}(t,r) = \frac {1}{2 (\alpha-1) r^{\alpha +l}}
\sum_{q=0}^{l-\alpha/2+1/2} (l-\alpha/2+1/2)^{\underline{q}}  \cdot
\frac {2^q \left(\alpha/2 \right)^{\overline{q}}}
{\alpha^{\overline{q}}} \, \mathcal{F} (t,r;\,2 \alpha - 1 +
q,\,l-1-q).
\end{equation}
Since $a(\eta)$ has compact support, it is advantageous to begin
with integration by parts
\begin{equation}
\int \limits_{-\infty}^{+\infty} d\eta \, \frac {P_l (\mu)}
{(\xi-\eta)^{\beta+L-k}} a^{(k)}(\eta)
 = \int \limits_{-\infty}^{+\infty} d\eta \, (-1)^k \frac {d^k} {d\eta^k} \left( \frac {P_l (\mu)} {(\xi-\eta)^{\beta+L-k}} \right)
 a(\eta)\,.\nonumber
\end{equation}
For $\mu$ as defined in (\ref{mu}) and for any function $g(\mu)$ the
following identity holds
\begin{equation}
\frac {d^k} {d\eta^k} \left( \frac {g(\mu)} {(\xi-\eta)^{\beta}}
\right) = \sum_{j=0}^k \left( \begin{array}{c} k \\ j
\end{array} \right) (\beta+k-1)^{\underline{k-j}} \left( \frac
{r^2-(t-\xi)^2} {r} \right)^j \frac {g^{(j)}(\mu)}
{(\xi-\eta)^{\beta+k+j}}\,,
\end{equation}
hence
\begin{equation}
\mathcal{F} (t,r;\,\beta,\,L) = - \frac {2^{\beta}}{4 r^{l+1}} \int
\limits_{-\infty}^{+\infty} d\eta \, a(\eta) \, \sum_{0\leq j\leq k
\leq L} (-1)^k \left( \begin{array}{c} k \\ j \end{array} \right)
c_{L,k} (\beta+L-1)^{\underline{k-j}}   \frac {1} {r^j} \int
\limits_{t-r}^{t+r} d\xi \, \frac {\left( r^2-(t-\xi)^2 \right)^j}
{(\xi-\eta)^{\beta+L+j}} P^{(j)}_{l} (\mu)\,. \label{master(2)}
\end{equation}
The sum over $k$ can be evaluated explicitly
\begin{equation}
\label{k-sum} \sum_{k=j}^L (-1)^k \left( \begin{array}{c} k
\\ j \end{array} \right) \frac {(2 L - k)!}
 {k! (L-k)!} (\beta+L-1)^{\underline{k-j}} = (-1)^L \left( \begin{array}{c} L \\ j \end{array} \right) (\beta-2)^{\underline{L-j}}\,.
\end{equation}
Let us define
\begin{equation}\label{I}
    \mathcal{I} := \frac {1} {r^j} \int \limits_{t-r}^{t+r} d\xi \,
\frac {\left( r^2-(t-\xi)^2 \right)^j} {(\xi-\eta)^{\beta+L+j}}
P^{(j)}_{l} (\mu) \,.
\end{equation}
Changing the integration variable from $\xi$ to $\mu$ and
integrating by parts, we get
\begin{equation}
\mathcal{I} = (-1)^j \frac {r^{j+1} (t-\eta)^{\beta-2+L-j}} {\left[
(t-\eta)^2-r^2 \right]^{\beta-1+L}} \int \limits_{-1}^{+1} d\mu \,
P_{l} (\mu) \frac {d^j} {d\mu^j} \left[ (1 - \mu^2)^j \left( 1 -
\frac {r} {t-\eta} \mu \right)^{\beta-2+L-j} \right]. \label{I}
\end{equation}
Using the identity \cite{wolfram}
\begin{equation}
\label{mu-to-k} \mu^k = \sum_{l=k,k-2,k-4,\dots} \frac {(2l+1) k!}
{2^{(k-l)/2} \left( \frac {k-l} {2} \right)! (k+l+1)!!}\, P_l
(\mu)\,,
\end{equation}
and expanding $\dfrac {d^j} {d\mu^j} \left[ (1 - \mu^2)^j \left( 1 -
\frac {r} {t-\eta} \mu \right)^{\beta-2+L-j} \right]$ in Taylor
series we get
\begin{eqnarray}
\mathcal{I} &=& (-1)^j \frac {r^{j+1} (t-\eta)^{\beta-2+L-j}}
{\left[ (t-\eta)^2-r^2 \right]^{\beta-1+L}} \,
\sum_{n=0}^{\beta-2+L} (j+n)^{\underline{j}} \, \int
\limits_{-1}^{+1} d\mu \, P_{l} (\mu) \mu^{n}
\nonumber\\
&\times&  \sum_{m=0}^{\lfloor (j+n)/2 \rfloor} \left(
\begin{array}{c} j \\ m \end{array} \right) \left( \begin{array}{c}
\beta-2+L-j \\ j+n-2m \end{array} \right) (-1)^{j+n+m} \left( \frac
{r} {t-\eta} \right)^{j+n-2m}
\nonumber\\
&=& \frac {r^{l+1} (t-\eta)^{\beta-2+L-l}} {\left[ (t-\eta)^2 - r^2
\right]^{\beta-1+L}} \,  \sum_{n=0}^{\lfloor (\beta-2+L-l) / 2
\rfloor} (j+l+2n)^{\underline{j}} \, \int \limits_{-1}^{+1} d\mu \,
P_{l} (\mu) \mu^{l+2n}
\nonumber\\
&\times&  \sum_{m=0}^{\lfloor (j+l+2n)/2 \rfloor} \left(
\begin{array}{c} j \\ m \end{array} \right) \left(
\begin{array}{c} \beta-2+L-j \\ j+l+2n-2m \end{array} \right)
(-1)^{l+m} \left( \frac {r} {t-\eta} \right)^{2j+2n-2m}
\nonumber\\
&=& \frac {r^{l+1} (t-\eta)^{\beta-2+L-l}} {\left[ (t-\eta)^2 - r^2
\right]^{\beta-1+L}} \,  \sum_{n=0}^{\lfloor (\beta-2+L-l) / 2
\rfloor} (j+l+2n)^{\underline{j}} \,  \, 2^{l+1} \frac {(l+2n)!
(l+n)!} {n! (2l+2n+1)!}
\nonumber\\
&\times&  \sum_{m=0}^{\lfloor (j+l+2n)/2 \rfloor} \left(
\begin{array}{c} j \\ m \end{array} \right) \left(
\begin{array}{c} \beta-2+L-j \\ j+l+2n-2m \end{array} \right)
(-1)^{l+m} \left( \frac {r} {t-\eta} \right)^{2j+2n-2m}
\label{I(2)}\,.
\end{eqnarray}
Collecting the results of (\ref{k-sum}, \ref{I}, \ref{I(2)}) and
plugging them into (\ref{master(2)}) we get
\begin{equation}
\mathcal{F} (t,r;\,\beta,\,L)
 = - \frac {2^{\beta+l+1}}{4} \int \limits_{-\infty}^{+\infty}
d\eta \, a(\eta) \, \frac {(t-\eta)^{\beta-2+L-l}} {\left[
(t-\eta)^2 - r^2 \right]^{\beta-1+L}} \sum_{n=0}^{\lfloor
(\beta-2+L-l) / 2 \rfloor} \frac {(l+2n)! (l+n)!} {n! (2l+2n+1)!}
(-1)^{L+l} L! \, S (\beta, L), \label{master(3)}
\end{equation}
where
\begin{equation}
\!\!S (\beta, L) = \sum_{j=0}^{L} \left( \begin{array}{c}
\beta-2 \\ L-j \end{array} \right) \left( \begin{array}{c} j+l+2n \\
j \end{array} \right)  \sum_{m=0}^{\lfloor (j+l+2n)/2 \rfloor}\!\!\!
(-1)^{m} \left( \begin{array}{c} j
\\ m \end{array} \right) \left( \begin{array}{c} \beta-2+L-j \\
j+l+2n-2m \end{array} \right) \left( \frac {r} {t-\eta}
\right)^{2j+2n-2m} \,.
\end{equation}
%%%%%%%%%%%%%%%%%%%%%%%%%%%%%%%%%%%%%%%%%%%%%%%%%%%%%%%%%%%%%%%%%%%%%%%%%%%%%%%%%%%%%%%%%%%%
%%%%%%%%%%%%%%%%%%%%%%%%%%%%%%%%%%%%%%%%%%%%%%%%%%%%%%%%%%%%%%%%%%%%%%%%%%%%%%%%%%%%%%%%%%%%
\subsection{First-order approximation}
%%%%%%%%%%%%%%%%%%%%%%%%%%%%%%%%%%%%%%%%%%%%%%%%%%%%%%%%%%%%%%%%%%%%%%%%%%%%%%%%%%%%%%%%%%%%
%%%%%%%%%%%%%%%%%%%%%%%%%%%%%%%%%%%%%%%%%%%%%%%%%%%%%%%%%%%%%%%%%%%%%%%%%%%%%%%%%%%%%%%%%%%%
To evaluate the first iterate $\phi_1(t,r)$ we apply the formula
(\ref{master(3)}) with $\beta=\alpha$ and $L=l$. Then
\begin{equation}
S(\alpha,l) = \sum_{j=0}^{l} \left( \begin{array}{c} \alpha-2
\\ l-j \end{array} \right) \left( \begin{array}{c} l+2n+j
\\ j \end{array} \right) \sum_{m=(j-l)/2}^{j+n} \!\!\!(-1)^{j+n-m} \left(
\begin{array}{c} j \\ j+n-m \end{array} \right) \left(
\begin{array}{c} \alpha-2+l-j \\ l-j+2m \end{array} \right) \left(
\frac {r} {t-\eta} \right)^{2m}\,, \label{SI}
\end{equation}
where we shifted the summation index $m\rightarrow j+n-m$. Next, we
interchange the order of summation according to
\begin{eqnarray}
&& [ 0 \leq j ] [ j \leq l ] [ m-n \leq j ] [ j \leq 2m+l ]
\nonumber\\
&\Leftrightarrow& [ -\frac {l}{2} \leq m < 0 ] [ 0 \leq j \leq l+2m
] \, + \, [ 0 \leq m < n ] [ 0 \leq j \leq l ] \, + \, [ n \leq m
\leq l+n ] [ m-n \leq j \leq l ]\,,\nonumber
\end{eqnarray}
and convert the sum over $j$ into a generalized hypergeometric
function \cite{gkp}. Defining
\[
t_j = (-1)^{j+n-m} \left( \begin{array}{c} \alpha-2 \\ l-j
\end{array} \right) \left( \begin{array}{c} l+2n+j \\ j \end{array}
\right) \left( \begin{array}{c} j \\ j+n-m \end{array} \right)
\left( \begin{array}{c} \alpha-2+l-j \\ l-j+2m \end{array}
\right)\,,
\]
we see that $t_0\neq 0$ iff $n=m$, thus the sums for $[ -\frac
{l}{2} \leq m < 0 ]$ and $[ 0 \leq m < n ]$ do not contribute to
(\ref{SI}) and we are left with
\begin{eqnarray}
 S (\alpha, l)&=& \sum_{m=n}^{n+l} \left( \frac {r}
{t-\eta} \right)^{2m} \sum_ {j=0}^{l+n-m} (-1)^{j} \left(
\begin{array}{c} \alpha-2 \\ l+n-m-j \end{array} \right) \left(
\begin{array}{c} l+n+m+j \\ j+m-n \end{array} \right)
\nonumber\\
& \times & \left( \begin{array}{c} j+m-n \\ j \end{array} \right)
\left(
\begin{array}{c} \alpha-2+l+n-m-j \\ l+n+m-j \end{array} \right),
\end{eqnarray}
where we  shifted the summation index $j \rightarrow j+m-n$.
Defining
\[
\tilde{t}_j = (-1)^{j} \left( \begin{array}{c} \alpha-2 \\ l+n-m-j
\end{array} \right) \left( \begin{array}{c} l+n+m+j \\ j+m-n
\end{array} \right) \left( \begin{array}{c} j+m-n \\ j \end{array}
\right) \left( \begin{array}{c} \alpha-2+l+n-m-j \\ l+n+m-j
\end{array} \right)
\]\,,
we see that
\begin{eqnarray}
\tilde{t}_0 &=& \frac {(\alpha-2)^{\underline{l+n-m}}} {(l+n-m)!}
\cdot \frac {(l+n+m)!} {(m-n)! (l+2n)!} \cdot \frac
{(\alpha-2+l+n-m)^{\underline{l+n+m}}} {(l+n+m)!}\nonumber
\end{eqnarray}
and
\[
\frac {\tilde{t}_{j+1}} {\tilde{t}_j} = \frac {(j - (l+n-m)) (j -
(l+n+m)) (j + (l+n+m+1)) } {(j + ((\alpha-1) - (l+n-m))) (j +
(-(\alpha-2) - (l+n-m)))
 (j +1)}\,,
\]
hence
\begin{eqnarray}
S (\alpha, l) &=& \sum_{m=n}^{n+l} \left( \frac {r} {t-\eta}
\right)^{2m} \frac {(\alpha-2)^{\underline{l+n-m}}} {(l+n-m)!} \cdot
\frac {(\alpha-2+l+n-m)^{\underline{l+n+m}}} {(m-n)! (l+2n)!}
\nonumber\\
&\times& F\left( \left. \begin{array}{c} - (l+n-m),\, - (l+n+m),\,
(l+n+m+1) \\ (\alpha-1) - (l+n-m),\, -(\alpha-2) - (l+n-m)
\end{array} \right| 1 \right)
\nonumber\\
&=& \sum_{m=n}^{n+l} \left( \frac {r} {t-\eta} \right)^{2m} 2^{1 +
2(l+n-m)} \pi \,\frac {(\alpha-2)^{\underline{l+n-m}}} {(l+n-m)!}
\cdot \frac {(\alpha-2+l+n-m)^{\underline{l+n+m}}} {(m-n)! (l+2n)!}
\nonumber\\
&\times& \frac {\Gamma(-(\alpha-2) - (l+n-m)) \Gamma((\alpha-1) -
(l+n-m))} {\Gamma \left( - \frac {\alpha-3} {2} + m \right) \Gamma
\left( - \frac {\alpha-2} {2} - (l+n)  \right) \Gamma \left( \frac
{\alpha} {2} + m \right) \Gamma \left( \frac {\alpha-1} {2} - (l+n)
\right)}\,, \label{SI(2)}
\end{eqnarray}
where in the last equation we used the identity
\[
F\left( \left. \begin{array}{c} a+1,\, -a,\, (b+c-1)/2 \\ b,\, c
\end{array} \right| 1 \right) = 2^{2-(b+c)} \pi \frac {\Gamma(b)
\Gamma(c)} {\Gamma \left( \frac {b-a} {2} \right) \Gamma \left(
\frac {c-a} {2} \right)
 \Gamma \left( \frac {1+b+a} {2} \right) \Gamma \left( \frac {1+c+a} {2}
 \right)}\,.
\]
Substituting
\begin{equation}
(\alpha-2)^{\underline{l+n-m}} \Gamma((\alpha-1) - (l+n-m)) =
\Gamma(\alpha-1),\nonumber
\end{equation}
and
\begin{equation}
(\alpha-2+l+n-m)^{\underline{l+n+m}} \Gamma(-(\alpha-2) - (l+n-m)) =
(-1)^{l+n+m} \Gamma(-\alpha + 2 + 2m)\nonumber
\end{equation}
into (\ref{SI(2)}) we get
\begin{eqnarray}
\!S (\alpha, l) \! &\!=\!&\! \sum_{m=n}^{n+l} \left( \frac {r}
{t-\eta} \right)^{2m}\! \!\frac {(-1)^{l+n+m} 2^{1 + 2(l+n-m)} \pi}
{(l+n-m)! (m-n)! (l+2n)!} \frac {\Gamma(\alpha-1) \Gamma(-\alpha + 2
+ 2m)} {\Gamma \left( \frac {\alpha} {2} + m \right)   \Gamma \left(
- \frac {\alpha-3} {2} + m \right) \Gamma \left( - \frac {\alpha-2}
{2} - (l+n)  \right) \Gamma \left( \frac {\alpha-1} {2} - (l+n)
\right)}\,. \nonumber \label{SI(3)}
\end{eqnarray}
The last equation can be still simplified due to the identity
\begin{equation}
\label{GammasId} \frac {\Gamma(\alpha -1) \Gamma(-\alpha + 2)}
{\Gamma \left( \frac {\alpha} {2} \right) \Gamma \left( - \frac
{\alpha-3} {2} \right) \Gamma \left( - \frac {\alpha-2} {2} - l
\right) \Gamma \left( \frac {\alpha-1} {2} - l \right)} = \frac
{(-1)^l} {2 \pi} \left( \frac {\alpha-3} {2} \right)^{\underline{l}}
\left( \frac {\alpha} {2} \right)^{\overline{l}}.
\end{equation}
We have
\begin{eqnarray}
\label{Gamma1} \Gamma(-\alpha + 2 + 2m) &=& (-\alpha +
2)^{\overline{2m}} \Gamma(-\alpha + 2),\nonumber\\
\label{Gamma2} \Gamma \left( - \frac {\alpha-3} {2} + m \right)& =&
\left( - \frac {\alpha-3} {2} \right)^{\overline{m}} \Gamma \left( -
\frac {\alpha-3} {2} \right),\nonumber\\
\label{Gamma3} \Gamma \left( - \frac {\alpha-2} {2} - l - n \right)
&=& \frac {\Gamma \left( - \frac {\alpha-2} {2} - l \right)} {\left(
- \frac {\alpha-2} {2} - l - 1 \right)^{\underline{n}}}\,,\nonumber \\
\label{Gamma4} \Gamma \left( \frac {\alpha-1} {2} - l - n \right)
&=& \frac {\Gamma \left( \frac {\alpha-1} {2} - l \right)} {\left(
\frac {\alpha-1} {2} - l - 1 \right)^{\underline{n}}}\,,\nonumber\\
\label{Gamma5} \Gamma \left( \frac {\alpha} {2} + m \right) &= &
\left( \frac {\alpha} {2} \right)^{\overline{m}} \Gamma \left( \frac
{\alpha} {2} \right),\nonumber
\end{eqnarray}
and
\begin{equation}
 \frac {(-\alpha + 2)^{\overline{2m}}} {\left( - \frac {\alpha-3}
{2} \right)^{\overline{m}}} = 2^{2m} \left( - \frac {\alpha} {2} + 1
\right)^{\overline{m}},\nonumber
\end{equation}
so finally
\begin{eqnarray}
S (\alpha, l) &=& \sum_{m=n}^{n+l} \left( \frac {r} {t-\eta}
\right)^{2m} \frac {(-1)^{n+m} 2^{2(l+n)}} {(l+n-m)! (m-n)! (l+2n)!}
\left( \frac {\alpha-3} {2} \right)^{\underline{l}} \left( \frac
{\alpha} {2} \right)^{\overline{l}}
\nonumber\\
&\times& \frac {\left( - \frac {\alpha} {2} + 1
\right)^{\overline{m}} \left( - \frac {\alpha-2} {2} - l - 1
\right)^{\underline{n}} \left( \frac {\alpha-1} {2} - l - 1
\right)^{\underline{n}}} {\left( \frac {\alpha} {2}
\right)^{\overline{m}}}\,. \label{SI(4)}
\end{eqnarray}
Plugging (\ref{SI(4)}) into (\ref{master(3)}) with $\beta=\alpha$
and $L=l$ we get the expression (\ref{tailf1}).
%%%%%%%%%%%%%%%%%%%%%%%%%%%%%%%%%%%%%%%%%%%%%%%%%%%%%%%%%%%%%%%%%%%%%%%%%%%%%%%%%%%%%%%%%%%%
\subsection{Second-order approximation}
%%%%%%%%%%%%%%%%%%%%%%%%%%%%%%%%%%%%%%%%%%%%%%%%%%%%%%%%%%%%%%%%%%%%%%%%%%%%%%%%%%%%%%%%%%%%
%%%%%%%%%%%%%%%%%%%%%%%%%%%%%%%%%%%%%%%%%%%%%%%%%%%%%%%%%%%%%%%%%%%%%%%%%%%%%%%%%%%%%%%%%%%%
The calculation in the second order ($\beta=2\alpha-1+q$ and
$L=l-1-q$) is only a slight modification of what we have already
done in the first order. Following the same steps which led us from
(\ref{SI}) to (\ref{SI(3)}) we get
\begin{eqnarray}
 S (\beta, L)
 &=& \sum_{m=n}^{n+L} \left( \frac {r} {t-\eta}
\right)^{2m} \frac {(-1)^{l+n+m} 2^{1 + 2(L+n-m)} \pi} {(L+n-m)!
(m-n)! (l+2n)!}
\nonumber\\
&\times& \frac {\Gamma(\beta-1) \Gamma(-\beta + 2 + l-L + 2m)}
{\Gamma \left( \frac {\beta} {2} + \frac{l-L}{2} + m \right) \Gamma
\left( - \frac {\beta-3} {2} + \frac{l-L}{2} + m \right) \Gamma
\left( - \frac {\beta-2} {2} - \left(\frac{l+L}{2}+n\right)
\right)\Gamma \left( \frac {\beta-1} {2} -
\left(\frac{l+L}{2}+n\right) \right)}\,. \label{j-sum(3)II}
\label{SII(3)}
\end{eqnarray}
The last equation can be simplified due to the identity
\begin{eqnarray}
&& \frac {\Gamma(\beta - 1) \Gamma(-\beta + 2 + l-L)} {\Gamma \left(
\frac {\beta} {2} + \frac{l-L}{2} \right) \Gamma \left( - \frac
{\beta-3} {2} + \frac{l-L}{2} \right) \Gamma \left( - \frac
{\beta-2} {2} - \frac{l+L}{2} \right) \Gamma \left( \frac {\beta-1}
{2} - \frac{l+L}{2} \right)}
\nonumber\\
&=& \frac {(-1)^l} {2 \pi} \left( \frac {\beta-3-(l-L)} {2}
\right)^{\underline{L}} \left( \frac {\beta + l-L} {2}
\right)^{\overline{L}} (\beta-2)^{\underline{l-L}}\,,
\end{eqnarray}
which for $L=l$  reduces to (\ref{GammasId}). We have
\begin{eqnarray}
\label{Gamma1II} \Gamma(-\beta + 2 + l-L + 2m) &=& (-\beta + 2 +
l-L)^{\overline{2m}} \,\Gamma(-\beta + 2 + l-L)\,,\nonumber\\
\label{Gamma2II} \Gamma \left( - \frac {\beta-3} {2} + \frac{l-L}{2}
+ m \right) &=& \left( - \frac {\beta-3} {2} + \frac{l-L}{2}
\right)^{\overline{m}} \Gamma \left( - \frac {\beta-3} {2} +
\frac{l-L}{2} \right)\,,\nonumber\\
\label{Gamma3II} \Gamma \left( - \frac {\beta-2} {2} - \frac{l+L}{2}
- n \right) &=& \frac {\Gamma \left( - \frac {\beta-2} {2} -
\frac{l+L}{2} \right)} {\left( - \frac {\beta-2} {2} - \frac{l+L}{2}
- 1 \right)^{\underline{n}}}\,,\nonumber\\
\label{Gamma4II} \Gamma \left( \frac {\beta-1} {2} - \frac{l+L}{2} -
n \right) &=& \frac {\Gamma \left( \frac {\beta-1} {2} -
\frac{l+L}{2} \right)} {\left( \frac {\beta-1} {2} - \frac{l+L}{2} -
1 \right)^{\underline{n}}}\,,\nonumber\\
\label{Gamma5II} \Gamma \left( \frac {\beta} {2} + \frac{l-L}{2} + m
\right) &=& \left( \frac {\beta} {2} + \frac{l-L}{2}
\right)^{\overline{m}} \Gamma \left( \frac {\beta} {2} +
\frac{l-L}{2} \right)\,,\nonumber
\end{eqnarray}
and
\begin{equation}
 \frac {(-\beta + 2 + l-L)^{\overline{2m}}} {\left( - \frac
{\beta-3} {2} + \frac{l-L}{2} \right)^{\overline{m}}} = 2^{2m}
\left( - \frac {\beta} {2} + \frac{l-L}{2} + 1
\right)^{\overline{m}},\nonumber
\end{equation}
hence
\begin{eqnarray}
S (\beta, L) &\!=\!& \sum_{m=n}^{n+L} \left( \frac {r} {t-\eta}
\right)^{2m} \frac {(-1)^{n+m} 2^{2(L+n)}} {(L+n-m)! (m-n)! (l+2n)!} \label{SII(4)}\\
&\times& \left( \frac {\beta-3-(l-L)} {2} \right)^{\underline{L}}
\left( \frac {\beta + l-L} {2} \right)^{\overline{L}}
(\beta-2)^{\underline{l-L}} \frac {\left( - \frac {\beta} {2} +
\frac{l-L}{2} + 1 \right)^{\overline{m}} \left( - \frac {\beta-2}
{2} - \frac{l+L}{2} - 1 \right)^{\underline{n}} \left( \frac
{\beta-1} {2} - \frac{l+L}{2} - 1 \right)^{\underline{n}}} {\left(
\frac {\beta} {2} + \frac{l-L}{2} \right)^{\overline{m}}}.\nonumber
\end{eqnarray}
Plugging (\ref{SII(4)}) into (\ref{master(3)}) we get
\begin{eqnarray}
&& \mathcal{F} (t,r;\,2\alpha-1+q,\,l-1-q) = (-1)^{q} \,\frac
{2^{2\alpha+3l-2-q}}{4} \left( \alpha - \frac {5} {2}
\right)^{\underline{l-1-q}} \left( \alpha - 2 + l
\right)^{\underline{l-1-q}} (2\alpha-3)^{\overline{1+q}}
\nonumber\\
&\times& \int \limits_{-\infty}^{+\infty} d\eta \, a(\eta) \, \frac
{(t-\eta)^{2\alpha-4}} {\left[ (t-\eta)^2 - r^2
\right]^{2\alpha-3+l}}
\\
&\times& \sum_{n=0}^{\alpha-2}\! (-1)^{n} \frac {2^{2n}(l+n)!} {n!
(2l+2n+1)!} \left( -\alpha+1-l \right)^{\underline{n}} \left( \alpha
- \frac {3} {2} - l + q \right)^{\underline{n}} \, \sum_{m=n}^{
n+l-1-q} \!(-1)^{m} \left(
\begin{array}{c} l-1-q \\ m-n \end{array} \right) \frac {\left(
-\alpha + 2 \right)^{\overline{m}}} {\left( \alpha + q
\right)^{\overline{m}}} \left( \frac {r} {t-\eta}
\right)^{2m}.\nonumber
\end{eqnarray}
Substituting this into (\ref{a5}) and expanding in $1/t$ we have
\begin{eqnarray}
\phi_{2}(t,r) &=& \frac {1} {2(\alpha-1)} \cdot \frac
{2^{2\alpha+2l-2}}{4 (2l+1)!!} \cdot \frac {1} {t^{2\alpha+2l-2}}
\left[ A + 2(\alpha+l-1) \frac{B}{t} \, + \mathcal{O} \left(
\frac{1}{t^2} \right) \right]
\nonumber\\
&\times& \left( \sum_{q=0}^{l-(\alpha-1)/2} (-1)^{q}
(l-p)^{\underline{q}} \,\, \frac {2^q \left( \alpha/2
\right)^{\overline{q}}} {\alpha^{\overline{q}}} \left( \alpha -
\frac {5} {2} \right)^{\underline{l-1-q}} \left( \alpha - 2 + l
\right)^{\underline{l-1-q}} (2\alpha-3)^{\overline{1+q}} \right),
\end{eqnarray}
with $A$ and $B$ defined  in (\ref{ab}). Converting the sum over $q$
into the generalized hypergeometric function we get (\ref{tail2}).


\begin{thebibliography}{10}

\bibitem{st} W. Strauss and K. Tsutaya, Discrete Cont. Dyn. Sys. \textbf{3}, 175 (1997).

\bibitem{sz} N. Szpak, arXiv:0708.1185 [math-ph]

\bibitem{ching} E. S. C. Ching et al., Phys. Rev. \textbf{D52}, 2118 (1995).

\bibitem{p} R. Price, Phys. Rev. \textbf{D5}, 2419 (1972).

\bibitem{l} E. W. Leaver, Phys. Rev. \textbf{D34}, 384 (1986).

\bibitem{gpp} C. Gundlach, R. Price and J. Pullin,  Phys. Rev. \textbf{D49}, 883
(1994).

\bibitem{b} L. Barack, Phys. Rev. \textbf{D59}, 044017 (1999).

\bibitem{dr} M. Dafermos and I. Rodnianski,  Invent. Math. \textbf{162}, 381
(2005).

\bibitem{car} V. Cardoso et al., Phys. Rev. \textbf{D68}, 061503 (2003).

\bibitem{k} J. G. Kingston, Quart. Appl. Math. \textbf{46}, 775 (1988).

\bibitem{ls} H. Lindblad and C. D. Sogge, Amer. J. Math. \textbf{118}, 1047 (1996).

\bibitem{hod} S. Hod, Class. Quantum Grav. \textbf{18}, 1311 (2001).

\bibitem{ik} A. Ishibashi and H. Kodama, Prog. Theor. Phys. \textbf{110}, 901 (2003).

\bibitem{gkp} R. L. Graham, D. E. Knuth and O. Patashnik, \textit{Concrete Mathematics} (Reading, Massachusetts: Addison-Wesley, 1994).

\bibitem{wolfram} http://mathworld.wolfram.com/

\end{thebibliography}
\end{document}